\documentclass[12pt]{extarticle}
\usepackage[mathscr]{euscript}
\usepackage{amsmath}
\usepackage[numbers]{natbib}
\usepackage{graphicx}
\usepackage[a4paper, margin=1in]{geometry}
\usepackage{xcolor}
\usepackage{longtable}

\title{Measuring the Change in European and US COVID-19 death rates}

\author{
Zeina S. Khan,\\
Frank Van Bussel,\\
\&\\
Fazle Hussain$^*$\\
Texas Tech University, Department of Mechanical Engineering\\ 
2703 7th Street, Box: 41021, Lubbock, TX 79409\\
Phone: 832-863-8364\\ 
$*$ fazle.hussain@ttu.edu\\}

\begin{document}
\maketitle
\begin{abstract}
By fitting a compartment ODE model for Covid-19 propagation to cumulative case and death data for US states and European countries, we find that the case mortality rate seems to have decreased by at least 80\% in most of the US and at least 90\% in most of Europe. These are much larger and faster changes than reported in empirical studies, such as the 18\% decrease in mortality found for the New York City hospital system from March to August 2020 \cite{horwitz2020trends}. Our reported decreases surprisingly do not have strong correlations to other model parameters (such as contact rate) or other standard state/national metrics such as population density, GDP, and median age. Almost all the decreases occurred between mid-April and mid-June, which unexpectedly corresponds to the time when many state and national lockdowns were released resulting in surges of new cases. Several plausible causes for this drop are examined, such as improvements in treatment, face mask wearing, a new virus strain, and potentially changing demographics of infected patients, but none are overwhelmingly convincing given the currently available evidence.
\end{abstract}

\section*{Introduction}
A novel strain of coronavirus, SARS-CoV-2, causing Covid-19 disease, was identified in December 2019 by Chinese Health authorities in the city of Wuhan (Hubei), China \cite{novelcoronavirus, world2020coronavirus22}. This disease has spread worldwide and many governments instituted measures to contain its outbreak, including city and state lockdowns and prohibiting travel from affected areas \cite{han2020lessons}. However, such restrictions are difficult to sustain in the long term, with millions of people being affected by poverty and unemployment \cite{han2020lessons}. As a result, many nations have eased population restrictions as of May 2020 to lessen the economic impact of the disease \cite{han2020lessons, hor2020hope, get2020as}. A global pandemic is ongoing, with over 40 million worldwide cases and 1.1 million deaths as of October 18, 2020 as declared by the World Health Organization (WHO) \cite{20201020-weekly-epi-update-10}. Currently, several European nations are considering reimposing lockdowns and mobility restrictions to contain the surging virus cases \cite{crow2020euro}.

Surprisingly, despite the large increases in Covid-19 cases in the United States and Europe since many countries and States eased lockdowns \cite{crow2020euro, merv2020it}, the number of deaths due to this virus has not mirrored the dramatically increased case counts. Though this trend has been noted by political and health commentators \cite{drum2020deathrate, justice2020deaths, campbell2020deathrate}, there are few mentions of any change of death rates in the epidemiological and modeling literature. Clinical observations of a continually decreasing death rate have been made in a New York City hospital system, with a rate that had dropped by 18.2 percent from March to August \cite{horwitz2020trends}. Corroborating this, time-series generated for a model-based study of Covid-19 in New York City imply that the infection-fatality risk dropped by approximately $\frac{1}{3}$ from early April to late May 2020 for people 65 years of age or older, whereas it barely changed for people less than 45 (fluctuations in the rate for the 45--64 years old group make the net effect difficult to ascertain) \cite{yang2020estimating}. Similarly, a review of English hospital surveillance data found that the survival of hospitalized Covid-19 patients in intensive care and high intensive units increased by approximately 11\% per week from late March up to the third week of June 2020 \cite{dennis2020improv}. This study notes that these improvements in survival were consistent across subgroups defined by age, ethnicity, and major co-morbidites, among others \cite{dennis2020improv}. While these observations are consistent with our own, possible underlying causes were not described in these reports. 

By fitting a compartmental ODE disease model to state and national case and death data we have been able to measure changes in the case mortality rate across entire jurisdictions for all US states plus Washington DC and Puerto Rico, and all European countries (except Russia) plus Turkey. Our finding is that in most of these jurisdictions the death rate for diagnosed individuals decreased dramatically ($\approx$ 80\% in the US and 90\% in Europe), and almost all jurisdictions had a decrease of at least 30\%. These decreases happened largely in late April, May, and early June as many jurisdictions were easing lockdowns which resulted in surging cases. Having checked several quantitative regional factors that could influence these fatality rates, including basic age demographics, population density, geographical location, and certain economic indicators, we have surprisingly not found strong correlations to the magnitude of the drop in death rate, or the initial or final death rates individually. Several plausible causes for this dramatic drop are examined, such as improvements in treatment, face mask wearing, a new virus strain, and potentially changing demographics of infected patients, but none alone convincingly explain the magnitude of change we have measured given the currently available evidence.

\section*{Calculating the Changes in Death Rate}
To calculate the change in death rate we used a slightly modified version of the compartment model first presented in \cite{khan2020predictive}. This is a SIR-based ODE model that includes extra compartments and transfer rates to deal with: detected versus undetected infecteds, isolation on diagnosis, effects of social distancing policies, and possible loss of immunity for recovered populations (both detected and undetected). As well, it uses a power-law incidence rate to adjust for the effect of heterogeneous population densities. While this model could have many uses, our original purpose was to try to measure what proportion of infecteds were eventually being detected (our finding: about half in almost all jurisdictions).

\begin{figure}
\centering
\includegraphics[width=0.5\linewidth]{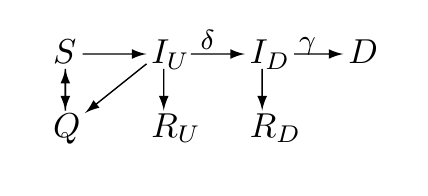}
\caption{Schematic of the compartments -- $S$ susceptible, $I_U$ undetected infected, $I_D$ detected infected, $D$ detected deceased, $Q$ sequestered, $R_U$ undetected recovered, and $R_D$ detected recovered. Transfer between compartments are indicated by arrows, where the detection rate $\delta$ and detected death rate $\gamma$ are highlighted. Note that $R_U$ and $R_D$ are also transferred back to $S$ at a particular rate due to loss of immunity. This is a small effect over the time scales we have considered here, therefore arrows were omitted for the sake of clarity.
}
\label{Fschema}
\end{figure}

As one can see in figure \ref{Fschema}, there is also a compartment for deaths of detected infected individuals. This was not necessary for the model per se; deaths are not part of the basic SIR model, which is static, and we did not include a compartment for deaths of undetected infecteds either. However, deaths due to COVID-19 are a readily available statistic, and possibly more trustworthy than caseloads. So on the assumption that the proportion of serious-enough-to-be-diagnosed cases which are fatal is relatively stable aspect of disease over the longer term, we added to detected deaths to the model to aid in its use for fitting to empirical data. Note: we were aware that case mortality rate in early days of any outbreak is often large, because of unfamiliarity of disease, and seriousness of first few diagnosed cases; but this tends to drop quickly, and the numbers of cases as a proportion of those eventually infected is small, so this transient effect did not affect our fits.

Coding of the ODE solutions and fitting routines were done in Matlab \cite{khan2020predictive}, with empirical data (cumulative cases and deaths, for the US by county, and globally by nation) obtained from Johns Hopkins University Center for Systems Science and Engineering \cite{csse2020data}. Since the model is non-linear, fitting requires an iterative search through the solution space, so there is no guarantee of obtaining optimal solutions within any set running time, but we found that with various adjustments to the fit parameters we were able to get good fits for all US states within a couple of hours (the criterion we use for goodness of fit was that the coefficient of determination $R^2 >= 0.95$). We were also able to fit all European countries Covid-19 case and death data (except Russia) using the same optimized code with similar results.

\begin{figure}
\centering
\includegraphics[width=0.75\linewidth]{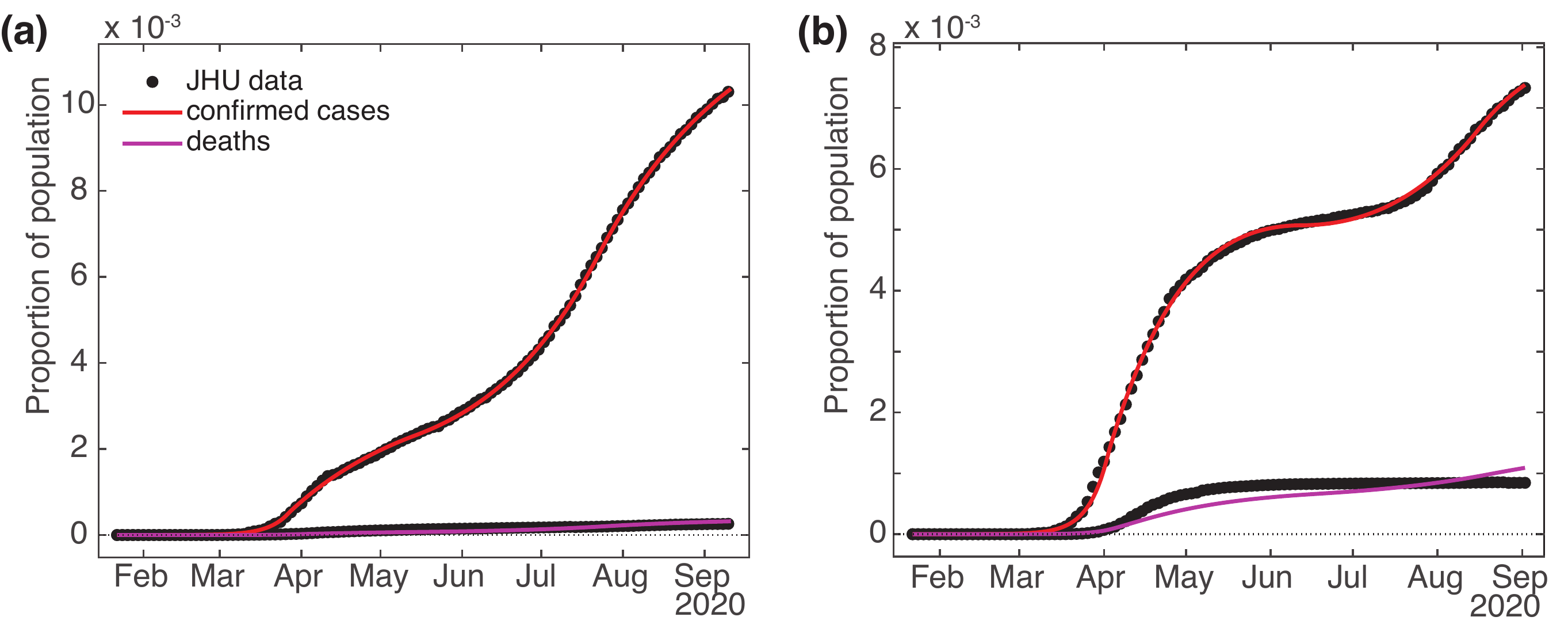}
\caption{Model fits using one death rate to cumulative case and death data for (a) Washington State and (b) Belgium. These two jurisdictions (which will be used to illustrative examples throughout the paper) were chosen at random from out of US and European datasets.}
\label{F1dfit}
\end{figure}

In the late summer of 2020 we started submitting predictions obtained from our model to the COVID-19 ForecastHub on GitHub [cite TBA]. When checking the results of 1-month-ahead death predictions, they seemed to be quite high, given that contemporary measures of deaths for various jurisdictions \cite{drum2020deathrate} were not showing spiking activity (though the recent new spikes in cases at the time suggested that deaths should be on the rise too). The problem was that the empirical (and therefore model) deaths were small in number compared with confirmed cases, so discrepancies between the model deaths and data were not readily apparent to visual inspection or the error criterion used by the fit routine, as can be seen from figure \ref{F1dfit}.

\begin{figure}
\centering
\includegraphics[width=0.75\linewidth]{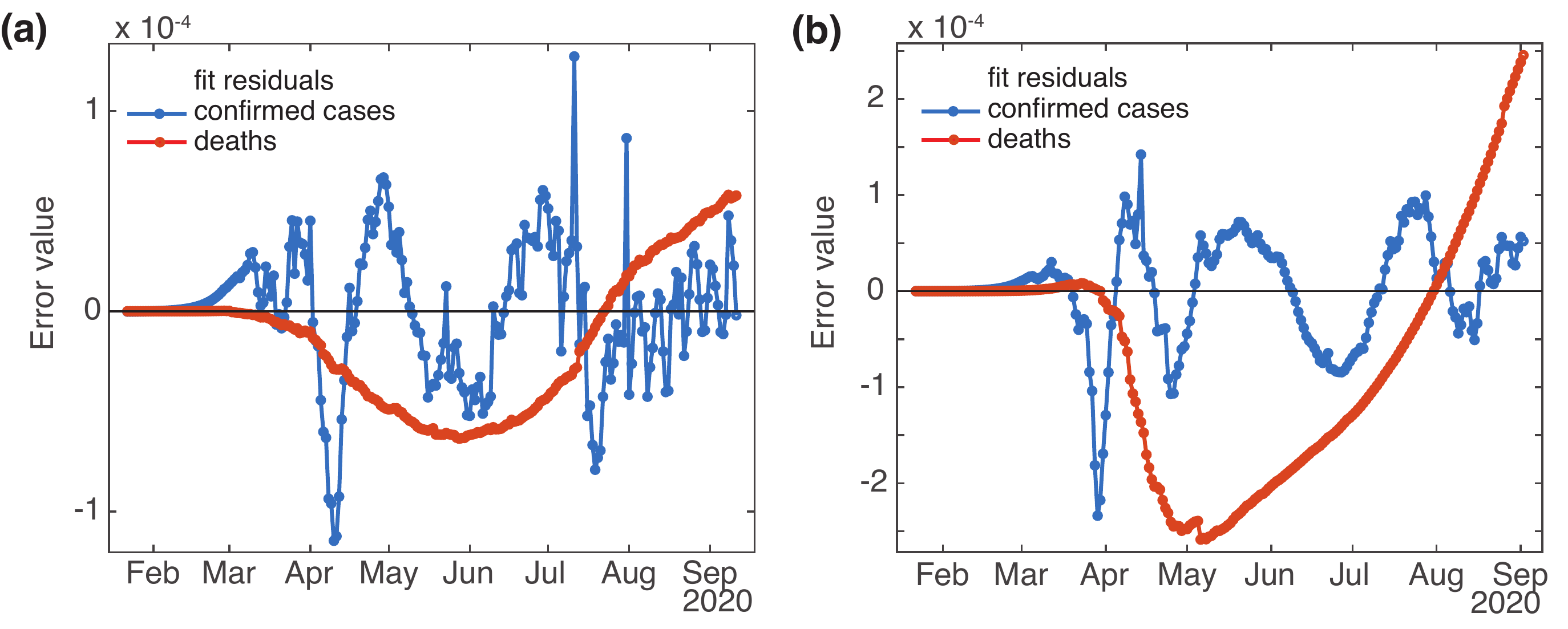}
\caption{One-death-rate model fit residuals for (a) Washington State and (b) Belgium. Note distinct time-dependent bias in error for fits to death data.}
\label{F1dres}
\end{figure}

\begin{figure}
\centering
\includegraphics[width=0.75\linewidth]{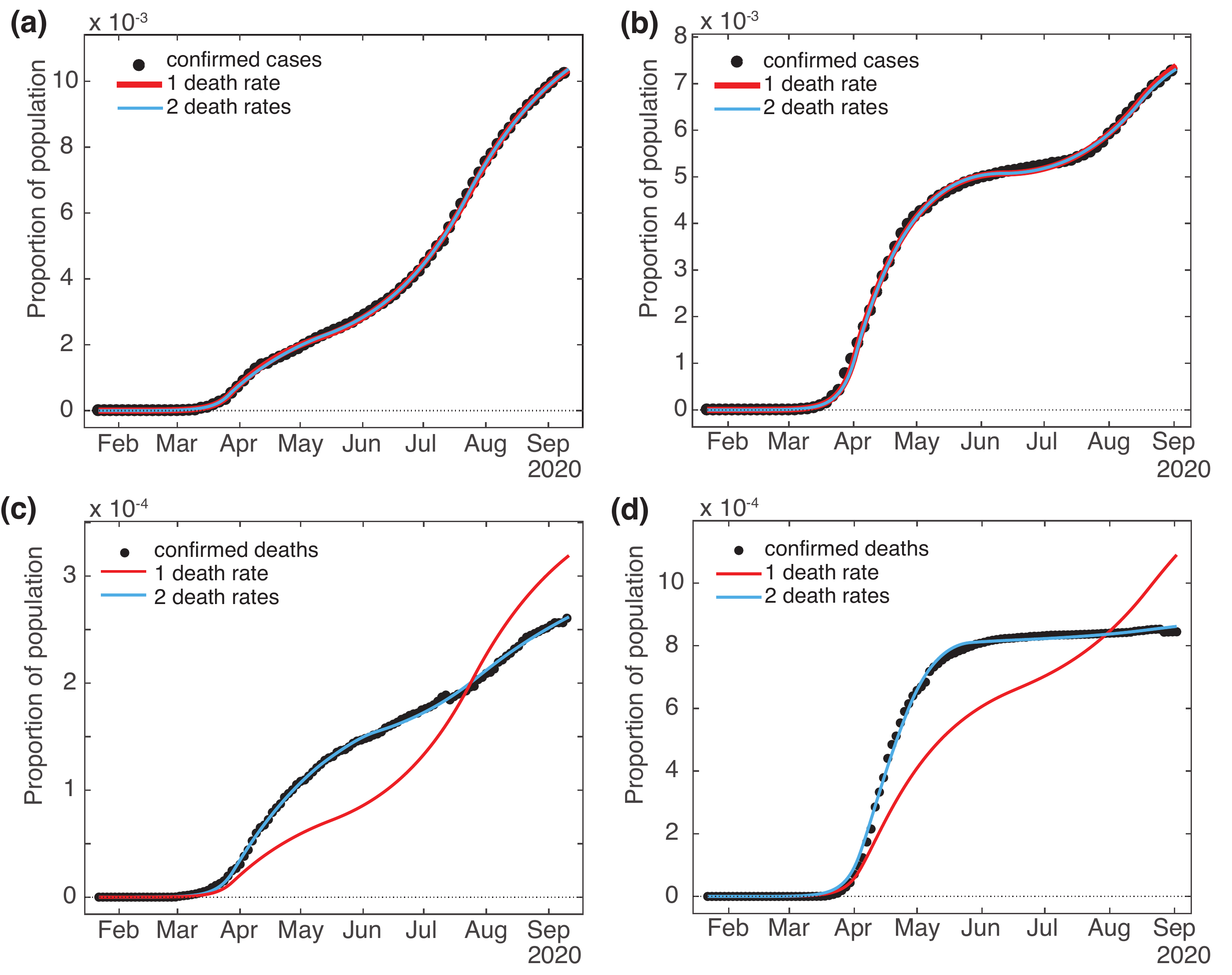}
\caption{Comparison of one-death-rate to two-death-rate model fits for (left) Washington State and (right) Belgium. (a and b) Cumulative confirmed cases, which show practically no change between versions. (c and d) Cumulative deaths.}
\label{F2dfit}
\end{figure}

However, a closer look at the death curves alone revealed that while the error level was within the desired tolerance, the residuals were not (more-or-less) randomly distributed across time, but showed a distinct bias, undershooting during the first half of the fit period and overshooting at later times, so the fit curve missed the contour actually described by the death data -- see figures \ref{F1dres} for residuals and \ref{F2dfit} (c) and (d) for closeups of the fits. This caused the slope of the model deaths at the end of the fit period to be greater than that implied by the empirical data, so that model projections of deaths into the fairly near future would overestimate the number significantly. Since this had not been a problem earlier, the implication was that the cumulative deaths were no longer shadowing the cumulative case count.
 
The solution was to make a simple modification of the model (and code) to incorporate a second death rate, with a changeover at a specific date. We then have two death rates, $\gamma_1$ and $\gamma_2$, plus a time parameter $t_{\gamma}$ specifying changeover day (from beginning of fit period). In the implementation of the ODE's, the rate changes linearly from $\gamma_1$ to $\gamma_2$ across 4 weeks centered on $t_{\gamma}$.  As well, in the fitter the death data is weighted to give it more emphasis in relation to the confirmed case data. All other aspects of the model and implementation were kept as is; i.e. all other rates (except for the already time-varying sequestration rate $q$) stay constant, and no change was made to the methodology with respect to solving the ODE's or non-linear minimization used by the fitter \cite{khan2020predictive}. The result is that the model confirmed case curves are practically identical to before, while the model death curve now falls quite precisely over the empirical data points -- see figure \ref{F2dfit}.

\begin{figure}
\centering
\includegraphics[width=0.75\linewidth]{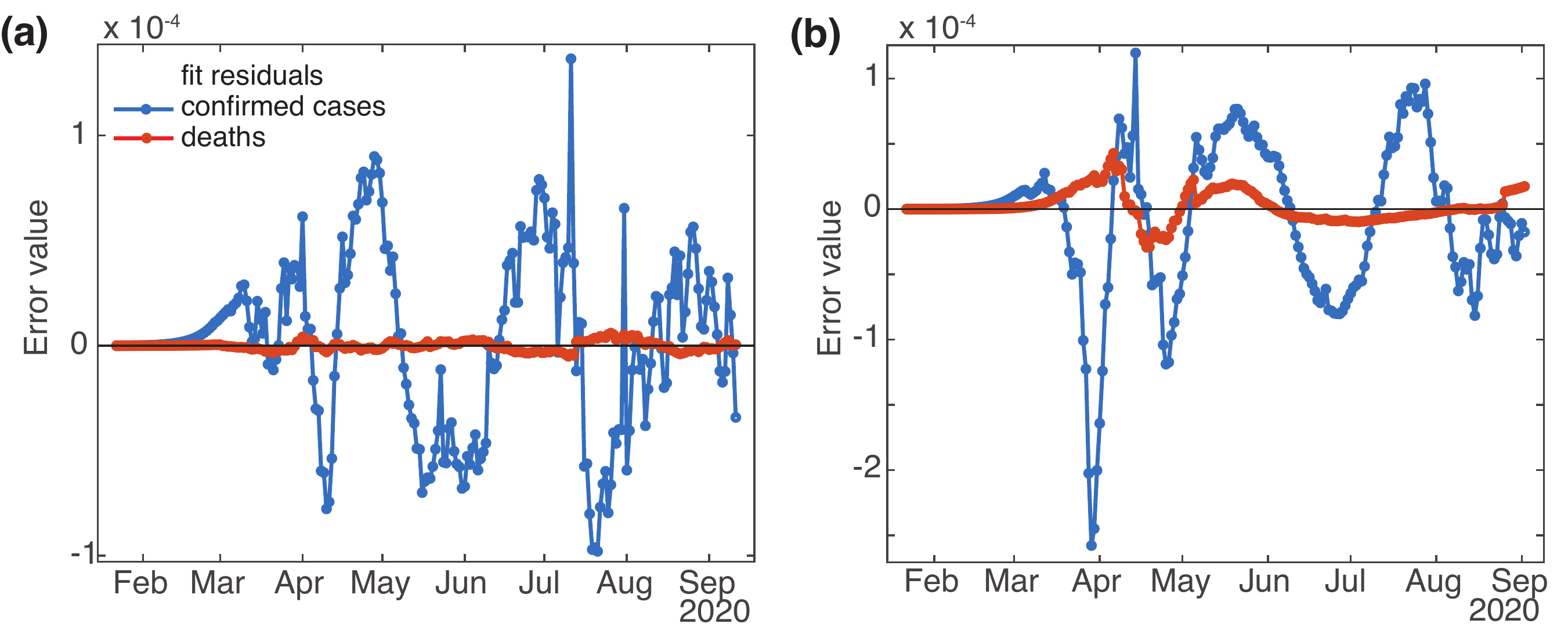}
\caption{Residuals for two-death-rate fits to (a) Washington state and (b) Belgium. The modified fit routine's differential weighting of confirmed case and death data results in much smaller residuals for death fit relative to confirmed case fit.}
\label{F2dres}
\end{figure}

Checking single death rate fits done at the beginning of September against empirical data going up to September 25, we find that deaths in both the US and Europe would be overestimated by approximately 50\% in both cases (a 98,000 overcount for the US, and a 118,000 overcount for Europe), while the two-rate version only misses by 9 and 2\% respectively.

\section*{Results}
\begin{figure}
\centering
\includegraphics[width=0.75\linewidth]{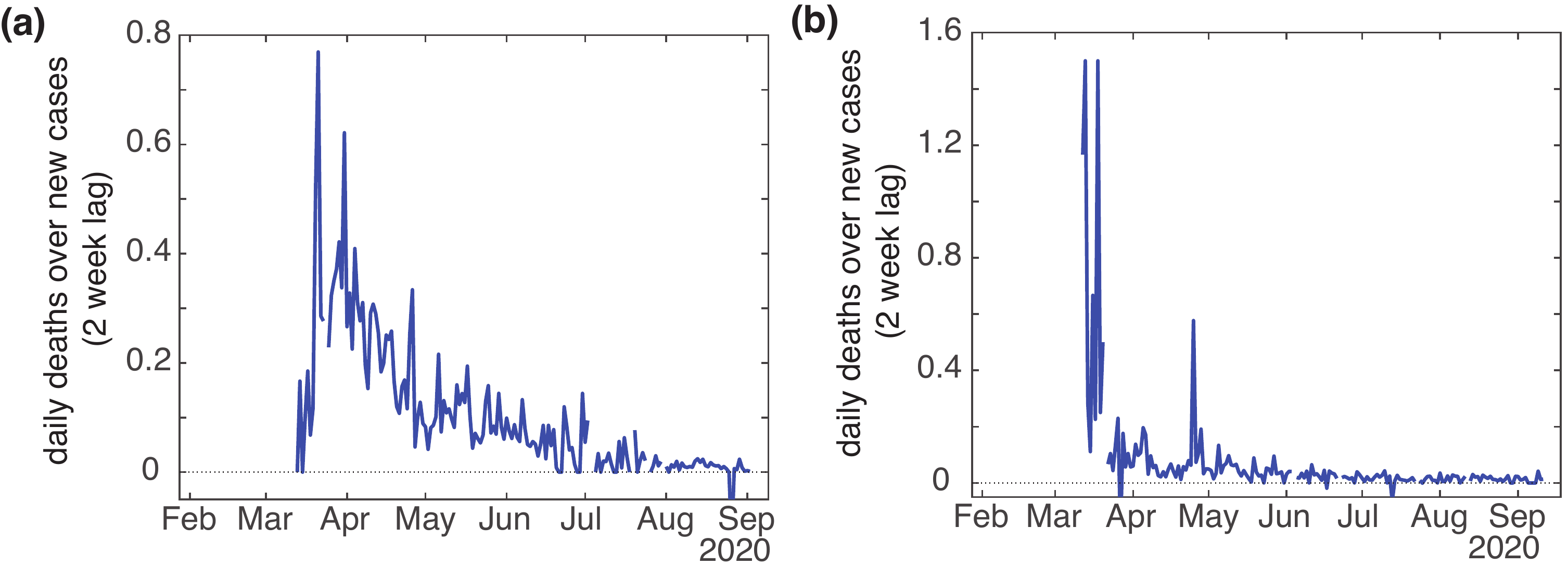}
\caption{Daily deaths divided by daily new cases with a two week delay for: (a) Washington state, and (b) Belgium.}
\label{Femp}
\end{figure}

To obtain our results model fits were done on 52 US jurisdictions (all states, plus Washington DC and Puerto Rico), and on 49 European zone countries (i.e. all Europe proper except for Russia, plus Turkey). The fit period was Jan 22 to Sept 2 for Europe and Jan 22 to Sept 11 for the US. Each fit provided the two death rates and a changeover time (in days from the start of the fit period, Jan 22 2020); percent change from $\gamma_1$ to rate $\gamma_2$ was calculated as well ($100 \times \frac{\gamma_2 - \gamma_1}{\gamma_1}$). See tables \ref{Tallus} and \ref{Talleu} in the Appendix for a full listing of rates, changeover day, and percent change for each US state and European country studied. Since the death rates apply only to detected infecteds, these can be seen as roughly equivalent to a very smoothed version of the case mortality rate, with weighting by current caseload -- see figure \ref{Femp}. Note that the empirical measures see a very high rate with a very large drop in the early days of the outbreak when only a handful of seriously ill people have been diagnosed, while any subsequent changes in the underlying rate are difficult to ascertain from the still fairly large day-to-day fluctuations.

\begin{figure}
\centering
\includegraphics[width=0.75\linewidth]{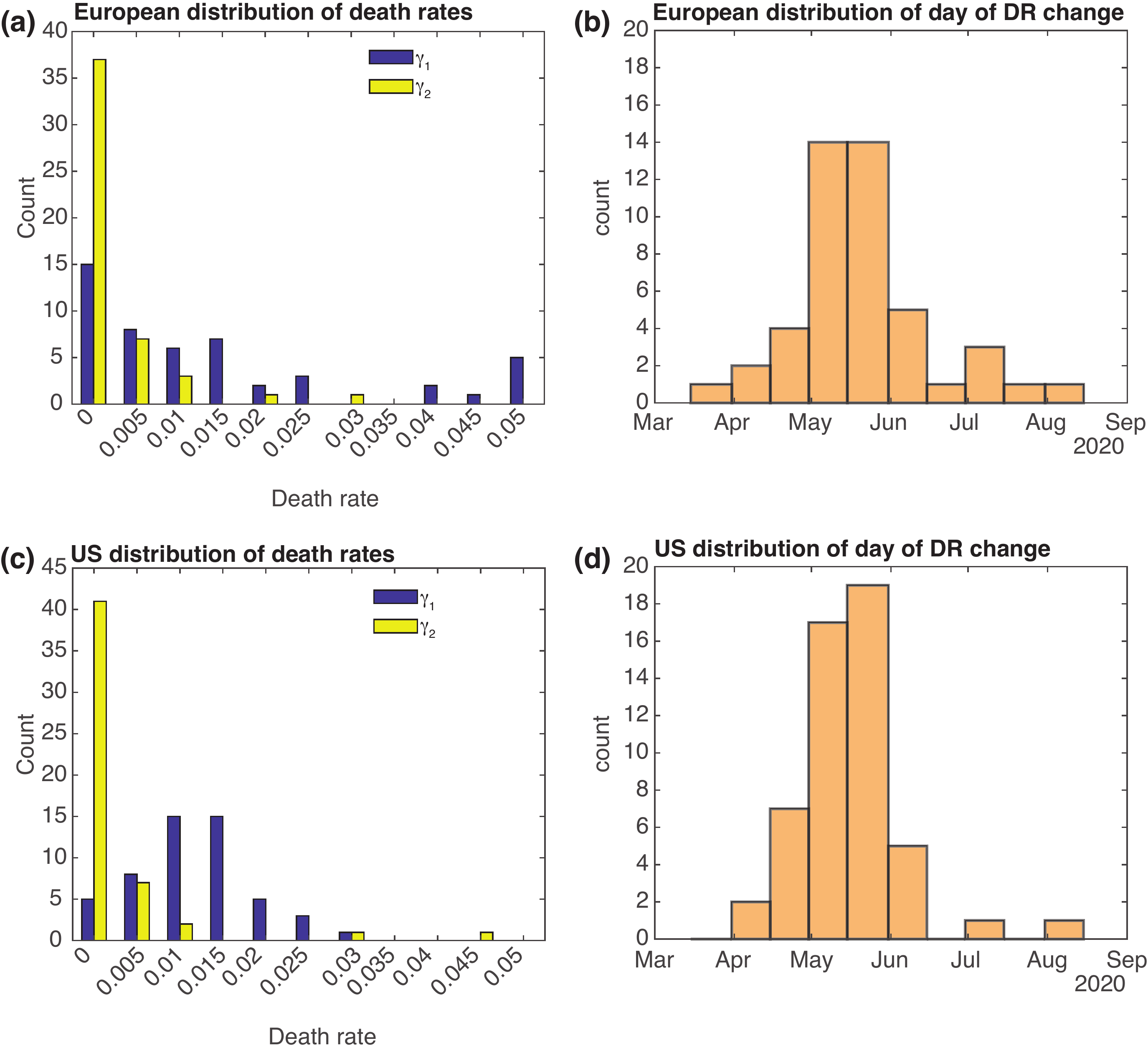}
\caption{
(a) Distribution of first $\gamma_1$ and second $\gamma_2$ death rates, and (b) distribution of the day of death rate change for all European countries (except Russia).
(c) Distribution of first $\gamma_1$ and second $\gamma_2$ death rates, and (d) distribution of the day of death rate change for all US states. 
}
\label{Fhist}
\end{figure}

We start with the rates themselves (see table \ref{Trate}). The initial death rate for detected infecteds is approximately 1\% in Europe and 1.5\% in the US, which is consistent with, if on low side of, values being hypothesized/calculated in March/April of this year \cite{owid2020mortality}. These go to approximately 1.5 per 1000 and 3 per 1000 respectively, a 5- or 6-fold drop. The changeover time in the fit period corresponds to the dates May 18 in Europe and May 15 in the US. As the standard deviation of the changeover time implies, most of the drops occurred within the period between mid-April and mid-June. Figure \ref{Fhist} shows how all the rates and changeover times are distributed. It is somewhat surprising that the second death rate $\gamma_2$ is much more narrowly distributed than the first death rate $\gamma_1$, and we have no explanation for this phenomenon.

\begin{table}
\begin{center}
\begin{tabular}{lll}
{\bf Metric} & {\bf Europe} & {\bf US} \\
\hline
Median $\gamma_1$ & 0.01058 & 0.014801 \\
Standard deviation $\gamma_1$ & 0.022353 & 0.0071521 \\
Median Changeover $t_{\gamma}$ (in days) & 117.7146 & 114.6639 \\
Standard deviation $t_{\gamma}$ (in days) & 33.1116 & 19.2085 \\
Median $\gamma_2$ & 0.0014119 & 0.0028296 \\
Standard deviation $\gamma_2$ & 0.0058485 & 0.0075316 \\
\hline
\end{tabular}
\caption{Statistics for the death rates $\gamma_1$ and $\gamma_2$, as well as the date of the change $t_{\gamma}$.}
\label{Trate}
\end{center}
\end{table}

\begin{figure}
\centering
\includegraphics[width=0.75\linewidth]{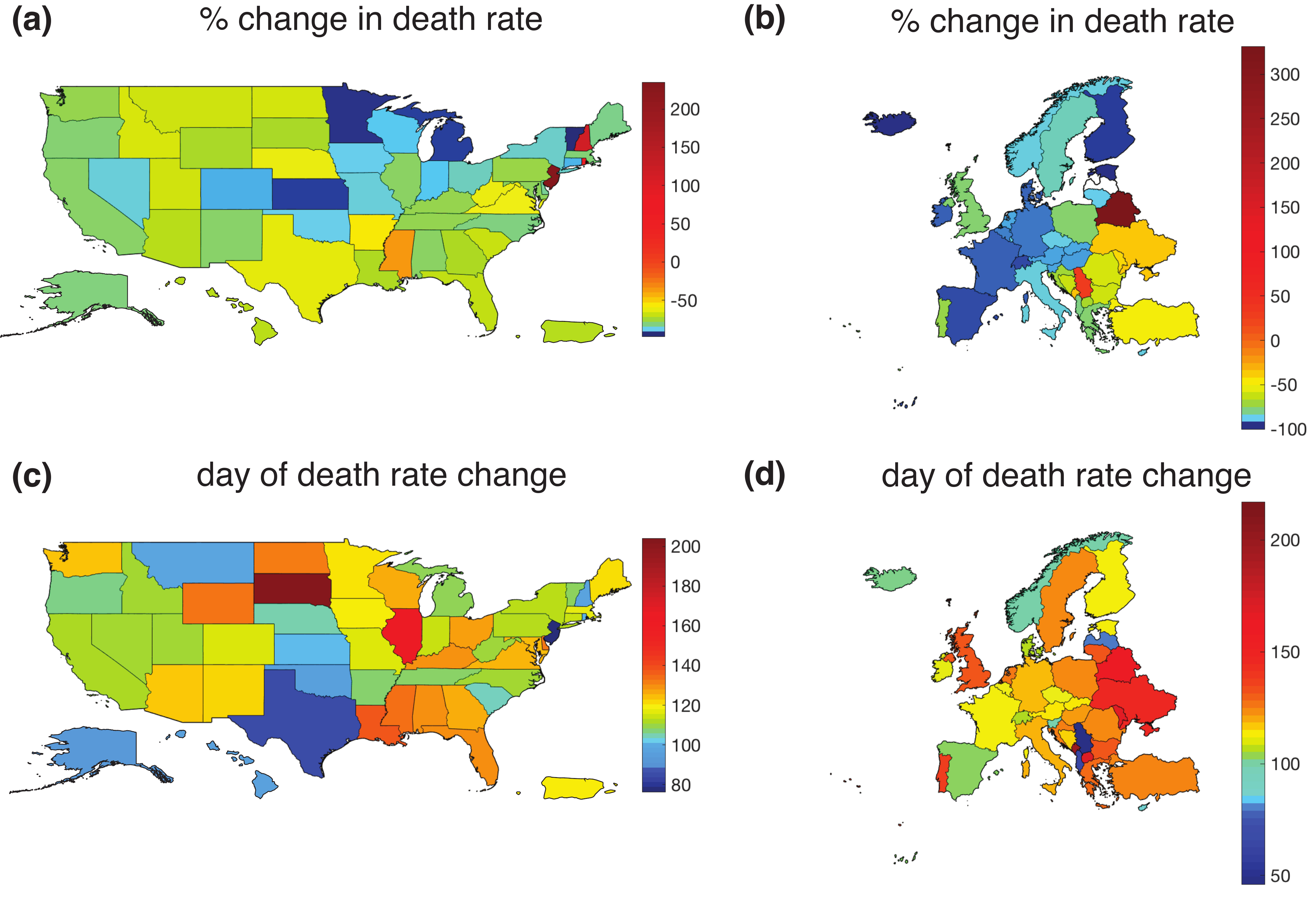}
\caption{
(a and b) Percent change of death rate for US and Europe. (c and d) changeover time (in days since January 22, 2020).
}
\label{Fmap}
\end{figure}

{\bf Change in death rate.} While it is to be expected that the case mortality rate of a disease will drift downward over time as medical treatments improve \cite{salazar2020treatment, klopfenstein2020tocilizumab, biran2020tocilizumab}, both the relatively tight timing and magnitude of the change in death rates are noteworthy; we see a decrease of approximately 90\% across Europe and 80\% across the US within a 2-month period. Table \ref{Tdrop} shows various statistics related to this drop (the value of the skewness measure most likely reflects the fact that -100\% is a sharp cutoff on the low end of the range of possible changes). Figure \ref{Fmap} gives maps of the US and Europe color-coded by the drop in rates and the changeover day; in the former particularly we see that the countries of western Europe for the most part saw large decreases, while eastern Europe is more variable. We also observe that US outliers with large positive changes in death rate are in the east. While there are clusters for the day of death rate change in Europe and the US, no clear pattern is apparent.

\begin{table}
\begin{center}
\resizebox{\linewidth}{!}{%
\begin{tabular}{lccl}
{\bf Metric} & {\bf Europe} & {\bf US} & {\bf comment} \\
\hline
\# Jurisdictions & 49 & 52 &  \\
Mean \% change in death rate & -70.0085 & -67.366 &  \\
Median \% change in death rate & -91.0148 & -80.6421 &  \\
Mode \% change in death rate & -94.8642 & -83.2356 &  \\
Outliers & 6 & 3 & with positive change \\
Greatest decrease & -100 & -97.3713 &  \\
Least decrease & -38.2119 & -38.0591 &  \\
Greatest increase & 330.9656 & 234.5559 & excluding countries 0 reported deaths \\
Standard deviation & 16.6705 & 11.6069 & (and below) excluding outliers \\
Skewness & 1.5267 & 0.96071 & $>$ 0 -- skews right \\
Kurtosis & 4.3511 & 4.5883 & $>$ 3 -- thicker tailed distribution than Gaussian \\
\hline
\end{tabular}
}
\end{center}
\caption{Statistics for the percent change in death rate.}
\label{Tdrop}
\end{table}

{\bf Outliers.} Not all jurisdictions saw decreases in death rates according to the measurement derived from out model. In the US three states, New Hampshire, New Jersey, and Rhode Island had increases, of 119, 235, and 55\% respectively. We note that New Jersey and Rhode Island had relatively late dates for the effective release from lockdown in comparison with other states, as measured by the model (June 16 and July 11 respectively). Mathematically, since these states did not open up at the same time as the others, their cases did not start rising dramatically again in the early summer, so the denominator  defining the case mortality rate stayed relatively low.

In Europe the outliers break down into two different groups. In the first case we have the Faroe Islands, Gibraltar, and Latvia, which had effectively no deaths in the period before the measured changeover (Faroe Islands and Gibraltar apparently had no deaths whatsoever during the entire fit period); in this case the astronomical positive changes in rate are merely an artifact of the extremely low initial rates given by the fitter. It should be noted that Latvia's neighbor Estonia had no recorded deaths in the period after changeover, and so achieved a 100\% drop; this suggests that the death statistics in the Baltic states may themselves be an issue.

The second group of European outliers -- Belarus (331\% increase), Kosovo (78\%), and Serbia (30\%), like the US outliers, are more perplexing. The latter two were of course famously involved in a violent conflict in the 1990's; all three are not members of the EU. Aside from that we can note that these countries had relatively late outbreaks (with first deaths recorded on March 22, March 29, and April 28 respectively), resulting in a later surge of cases and deaths.

{\bf Correlations.} One may ask if there is a relation between the measured changes in death rates and various other metrics. However, with one rather trivial exception (to be discussed below) we found no strong correlations of the drop to either model-related quantities or a number of readily available state/national statistics; though admittedly, our search through national databases was not exhaustive. All correlations discussed below were calculated using Matlab's {\tt corrcoef} function, which gives the Pearson correlation coefficient.

The first place to look is within the model parameters themselves, and quantities derived from either the raw data or projections based on the fits. Across multiple fits we would expect some rates to move in tandem or opposition to others, and indeed, for both Europe and the US we see that the SIR-based contact rate has a strong negative correlation ($< -0.9$) with both the recovery rate for undetected infecteds and the detection rate, as well as a slightly weaker positive correlation ($> 0.67$) with the severity of the first social distancing intervention. In fact, one model parameter does correlate strongly with the drop in the death rate: the second death rate itself (0.72 for Europe, 0.97 US), which is hardly surprising. However, no other rates or data derived quantities had an absolute correlation $> 0.5$ for either the US or Europe, and only a scattering had the absolute correlation $> 0.33$; these latter all had different signs for the European and US fits, indicating that the relation was not particularly robust despite the magnitude. Only three model related quantities other than the second death rate had absolute correlations $\geq 0.1$ with same sign for the US and Europe: loss of immunity rate (negative), initial condition (proportion of population infected on day 1 of fit period, negative), and proportion of unknown recovereds on last day of fit period (positive). In all these cases the absolute correlation was $<$ 0.22, so rather weak. 

\begin{table}
\begin{centering}
\begin{tabular}{lclcl}
{\bf Metric} & {\bf Europe} & {\bf US} \\
\hline
Population & -0.03319 & -0.045653 \\
Area & 0.057402 & -0.15576 \\
Pop. density & -0.080492 & 0.015033 \\
Latitude & -0.02293 & 0.062915 \\
Longitude & 0.33451 & 0.16866 \\
GDP \cite{USGDP2020, EUGDP2020} & -0.16385 & -0.05716 \\
GDP per capita \cite{USGDP2020, EUGDP2020} & -0.33194 & -0.07874 \\
Gini coefficient \cite{USGini2020, EUGini2020} & -0.12044 & 0.10396 \\
Median age \cite{USage2020, EUage2020} & -0.25117 & 0.19846 \\
\hline
\end{tabular}
\caption{Pearson correlation coefficients for \% changes in death rates and state/national statistics. Population, area, population density, latitude, and longitude data were obtained from Johns Hopkins University alongside Covid-19 data \cite{csse2020data}. }
\label{Tcorr}
\end{centering}
\end{table}

Since the correlations to standard state/national statistics may be of more general interest, these are given in table \ref{Tcorr}. As with the model parameters, most correlations here are quite weak and have different signs between Europe and the US; only longitude has non-trivial (though not strong) correlations of the same sign, which is apparent from figure \ref{Fmap}(a) showing consistently larger drops in percent death rate change in Western Europe than Eastern Europe. As mentioned above, we did not check many other possible quantities (eg. educational attainment, per capita health care expenditures, etc.) since each requires finding and converting new data extraneous to our main project; in particular, certain epidemiological data, such as COVID-19 testing rate (which is itself time-varying), might yield interesting results with more intensive comparison techniques. Note that correlations between the individual death rates and changeover day with the other model parameters and state/national statistics were also calculated, and as well did not show any strong or surprising correlations (data not shown).

\begin{figure}
\centering
\includegraphics[width=0.75\linewidth]{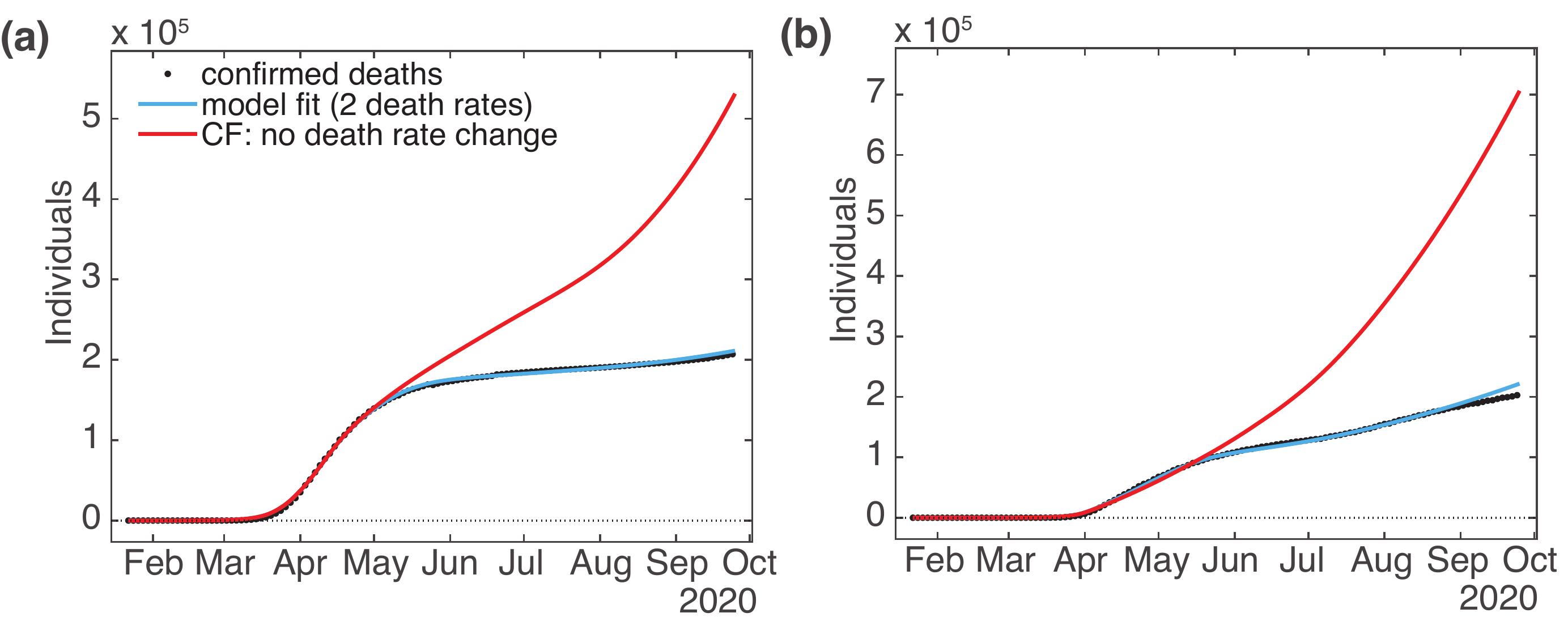}
\caption{Model fit to confirmed Covid-19 deaths, and the number of deaths predicted for the counterfactual (CF) scenario of no change in death rates for: (a) Europe, and (b) the United States.
}
\label{Fcfdeaths}
\end{figure}

{\bf Counterfactual Scenario.} Our implementation allowed us to run counterfactual simulations to test various suppositions by rerunning the ODE solver on the model with changed parameter values. By suppressing the second death rate, we are able to estimate what the deaths outcome would be if no change in rate had occurred. Figure \ref{Fcfdeaths} shows plots of deaths data, model fits, and counterfactual projections for Europe and the US. As one would expect, if the rate had not changed the number of deaths by September 25 would have been much greater, more than triple in the US (from $\approx$ 204,000 to 706,000) and more than double in Europe (from $\approx$ 208,000 to 531,000). Since the effect on the cumulative confirmed cases was minimal, we have not shown these plots.

\section*{Discussion and Conclusions}
There are several factors expected to affect fatality rates over the course of a pandemic. Improvements in medical treatments are to be expected as knowledge about the disease increases. For example, aggregated data suggests that transfusing (high anti-RBD IgG titer) convalescent plasma early in the hospital course of Covid-19 patients significantly reduces mortality by approximately 6\% in comparison with control patients \cite{salazar2020treatment}. Additional independent studies have shown that administering tocilizumab (a recombinant monoclonal antibody that can mitigate cytokine release syndrome) to patients admitted to intensive care with Covid-19 have a 23\% \cite{klopfenstein2020tocilizumab} and a 12\% \cite{biran2020tocilizumab} reduction in mortality, compared with patients receiving standard care. Importantly, a clinical outcomes study reported that patients who presented in hospital with sufficient vitamin D levels ($\geq$ 30 ng/ml) had reduced mortality rates by 10\% in comparison with Covid-19 patients with insufficient ($<$ 30 ng/ml) vitamin D \cite{maghbooli2020vitamin}, which suggests that lowly toxic supplementation and increased sun exposure can affect a population's outcome. The studies above also suggest that improvements in Covid-19 treatments since the start of the pandemic can reduce a population's overall mortality by at least 20\%, which is a smaller factor than we have measured.

It has also been suggested that mask wearing can reduce the mortality rate of Covid-19 via two different means. First, a face mask worn by an infected person forms a barrier for transmission of respiratory droplets to susceptible populations, thus reducing transmissibility of the disease \cite{betsch2020social, haischer2020wearing}. It may be reasonable to expect that populations with widespread mask usage and clear government guidelines may have a reduction in contact rate $\beta$ associated with policy implementation if the policy had been implemented after occurrence of exponential growth in cases \cite{titus2020mathematical}. We have not observed the need for such a reduction to fit cumulative cases in any country or state well. In any case, while reduced case counts (if we had seen them) would result in fewer deaths overall, this says nothing about deaths per case. But it is also possible that wearing a face mask protects the wearer by reducing the SARS-CoV-2 inoculum that they are exposed to by infected people \cite{stvrivzova2020can}. Exposure to a low viral load may result in a less severe, possibly asymptomatic, infection with a lower chance of fatality \cite{gandhi2020masks}. So it is still possible that the changed Covid-19 death rates were have observed result from face mask wearing; the YouGov online survey reporting tool demonstrates that self-reported face mask wearing in public spaces in some European countries (Italy, Spain, France, and Germany)  rapidly increased to 80\% of the population or more between late March and May 2020 \cite{yougov2020masks}. A similar trend is observed in the United States, where self-reported face mask wearing in public places rose to 69\% at the end of May 2020 \cite{yougov2020masks}. However, self-reported face mask wearing in the Nordic nations of Finland, Denmark, Norway, and Sweden did not exceed 20\% of the population over the same time frame, and these nations also experienced very large drops in death rate. This evidence strongly suggests that if wearing face masks is a factor that affects the death rate change, it is not the only one. 

It is also possible for a virus to acquire mutations that alter its infectivity and lethality over time. Genomic analyses have demonstrated that the spike protein of SARS-CoV-2 has undergone an amino acid change from aspartic acid (referred to as D614) to glycine (referred to as G614) in the carboxy-terminal region of the S1 domain \cite{leung2020empirical, korber2020spike, gomez2020mapping}. The very rapid spread of the G614 mutant throughout Europe and parts of the Americas, monitored by Covid-19 genetic surveillance studies over time, suggests that it could be more transmissible \cite{leung2020empirical, korber2020spike, gomez2020mapping, mohammad2020higher}. One regional study conducted within a Houston hospital system showed that the virus strains originally introduced into the city in March 2020 were diverse, with both D614 and G614 types represented, however sequences taken during the much larger second wave that occurred in June 2020 were nearly all of the G614 type \cite{long2020molecular}. They found that patients with the G614 strains had higher nasopharyngial viral loads upon diagnosis; however, the authors did not find evidence connecting virus genotype with altered virulence \cite{long2020molecular}. Interestingly, a data correlation study found that the G614 mutation corresponds to higher case fatality rates in several countries \cite{becerra2020sars}. Given the available evidence, it seems likely that the highly prevalent G614 mutation is not less deadly than previous strains, which leaves the distinct possibility that there is a newer, less deadly mutation circulating.

Increasing testing can also significantly impact the case fatality rate of a disease, since detecting increasing numbers of cases will increase the denominator of the case fatality rate, and possibly lead to earlier detection of a disease leading to earlier treatment thereby also reducing mortality \cite{pachetti2020impact, leffler2020association}. While we are not aware of any studies examining correlations between the number of Covid-19 tests in time and case fatality rates, several studies examining regional differences in fatality and testing have occurred. One study comparing USA, Italy, UK, France, Spain, Sweden, and Germany found that case fatality rates, normalized by the ratio of tests to total number of positive cases, tended to cluster suggesting a correlation between mortality and testing rate \cite{pachetti2020impact}. A multivariable statistical study of Covid-19 mortality in 196 countries found that a 10 times decrease in per-capita testing was correlated with a 26\% increase in per-capita mortality, though this correlation was not found to be statistically significant \cite{leffler2020association}. Another statistical comparison of testing rates and mortality across French region borders found that performing an additional 1000 tests would save one life \cite{terriau2020impact}. Data available from the Johns Hopkins University Coronavirus Resource Center \cite{jhu2020testing} shows that US tests increased by approximately 12 times (from 0.1 to 1.2 million) from April through November 2020, suggesting that increased testing may have played some role in the large death rate decrease we have observed in nearly all US states and European countries.  

It is also possible that the age demographics of people more recently afflicted with Covid-19 have affected the mortality rate -- particularly if more young people than elderly have become infected -- who tend to be much less likely to have severe disease \cite{promislow2020gero}. Indeed, an analysis of Covid-19 cases that occurred worldwide between February and July, 2020 revealed that the number of infected people 15-24 years old increased from 5\% to 15\%. Cases of Covid-19 in the USA in people 18-22 years old increased by 55\% from August 2-Sept 5, 2020, and was highest among people between 20 and 29 years old, with more than 20\% of the total cases, in contrast with March 2020 where Covid-19 incidence was highest in people with ages over and including 60 years \cite{venkatesan2020changing}. In conjunction with this trend, some clinical reports indicate that Covid-19 has become less deadly across all age groups. It was reported that the mortality rate, adjusted for changes in demographics, had dropped by 18\% in a New York city hospital system from March to August 2020 \cite{horwitz2020trends}. Similarly, English hospital surveillance data found that the survival of Covid-19 patients in both intensive care and high intensive units increased by approximately 11\% per week from late March through June 2020 across age, ethnicity, and major co-morbidity subgroups \cite{dennis2020improv}. Given these observations, it appears that changes in age demographics of Covid-19 incidence do not fully explain our observed change in mortality over time. 

Lastly, we look at the possibility that the drop could be a statistical artifact caused by changes in the way death data is recorded and collected. It should be noted, that we (along with \cite{drum2020deathrate}) first noticed the change of death rate not as a drop in daily deaths versus total population, but as persistence of the previous trend when surges in the number of cases versus total population occurred after releases of lockdowns, where concomitant surges in deaths were to be expected.

Data revision is common for many publicly maintained statistics, not only in medical areas but also economics and demographics, since later figures often improve or correct earlier ones, which may be based partly on estimates or incomplete surveys. With respect to diseases or mortality, large upward revisions often gain public attention, since the implication is of prior negligence or coverup. During the current Covid-19 pandemic a couple of instances do leap out: China's April revision upward by 1290 deaths (which increased their then case mortality by 50\%) \cite{griff2020wuhan}, or Argentina's massive correction at the beginning of October \cite{staff2020update}.

There are legitimate reasons for changes in procedure that result in lower death counts and subsequent downward revisions. Many jurisdictions initially logged all deaths of Covid-19 infected individuals as deaths by Covid, presumably because in the early days of the pandemic the exact range of co-morbidities had not been determined; when later information is available to limit that range, non-Covid deaths of Covid-infected individuals can be placed in the appropriate category. This is the case for the UK revision in August. Previously, the UK had been counting all deaths of Covid-infected people within 60 days as death by Covid-19, which was reduced to 28 days; applied retroactively, this had the effect of reducing the UK Covid-19 death count by 5,377 ($\approx$ 13\% at the time) \cite{hene2020public}. Similarly, Washington State, which had been counting all deaths of anyone who tested positive at any point as Covid-19 deaths, officially adapted a more stringent protocol in mid-June, only listing a death as Covid-related if it was a specific factor mentioned in the death certificate \cite{nix2020news}. Case and death reductions may also occur for other reasons. In Belgium a downward revision, ostensibly to correct for double-counting in nursing homes, made news because it seemed to be timed to avoid the milestone of 10,000 Covid-19 deaths \cite{blenk2020belg}.

Downward revisions of past death statistics, if integrated properly into time-series data, should not have an adverse affect on any attempt to determine changes in case-mortality over time, whether by our model or other techniques. Our primary data source, the JHU CSSE Covid team \cite{csse2020data}, seems to have made every effort to revise past data to reflect current knowledge and practice. To begin with, they cross-reference many sources of their own, including the World Health Organization, the European Centre for Disease Prevention and Control (ECDC), the US Center for Disease Control, many other national health organizations (such as the National Health Commission of the People's Republic of China, Australia Government Department of Health, Italian Ministry of Health, etc.), practically all US state Departments of Health, many municipalities and US counties, news organizations such as the Los Angeles Times and BNO News, and even a few other Covid-19 tracking projects (presumably for confirmation) such as ``the COVID Tracking Project'' maintained by The Atlantic (https://covidtracking.com/data) and WorldoMeters Covid page (https://www.worldometers.info/coronavirus/).

Importantly, when possible the JHU CSSE Covid team back-distribute revisions of past data (i.e. incorporate them on appropriate days in their currently available time series). According to their records, there have been 22 data modifications for European nations and 19 for US jurisdictions (which are tallied by county). As well, several large-scale back distributions have been done (twice for both New York City and Massachusetts; and once for the United Kingdom, Michigan, New Jersey, North Carolina, and Harris County, Texas). In general, such back distribution (whether an up or down revision) should make death data before mid-May more trustworthy rather than less.

An issue arises if jurisdictions adopt new protocols without revising past statistics, or do the revisions without back-distributing into the past data sets. In the JHU CSSE time series we used 36 US states and 21 European countries had decreases in cumulative deaths on 121 separate occasions, mostly by 1 or 2 cases. Since any decrease in cumulative deaths is a physical impossibility, the ones we see here presumably indicate data revisions which could not be back-distributed. For example, the time-series for Washington State has occasional negative day-to-day changes in death counts starting from mid June (when they changed their protocol) and lasting through July (when they seem to have finished whatever revisions they needed to make). The total number of deaths involved in these post hoc revisions is 2,463 for the European nations and 666 for the US states; while not trivial, these values could hardly account for the drops we have seen in the death rates detailed above. To determine how many downward non-back-distributed revisions occurred which did not result in negative day-to-day changes in cumulative deaths, or which countries, states, or counties quietly adopted different protocols or definitions without attempting to revise past totals, would require greater access to jurisdictional health agency revision and policy data than we have.

In conclusion, we have found that the case mortality rate of Covid-19 has dramatically decreased between mid-April and mid-June 2020 in many European countries and US states. While there are many plausible factors, such as improved medical technique, mask wearing, increased testing, viral mutation, demographics, or changes in recording of cases, that may have caused this, at this point we cannot conclusively say which, if any, are {\em the} cause, or if it is a combination of these or other subtle factors. This surprising finding warrants further attention.

\section*{Data Availability Statement}
Data for cumulative confirmed cases and deaths were obtained from the Johns Hopkins University (JHU) Center for Systems Science and Engineering, posted on the  GitHub  website \cite{csse2020data}. 

\section*{Conflicts of Interest}
Conflicts of Interest: None.

\section*{Acknowledgements}
This study was supported by TTU President's Distinguished Chair Funds. 


\section{Appendix}
\small
\begin{center}
\begin{longtable}{llllll}
\caption{\normalsize Death rates $\gamma_1$ and $\gamma_2$, day of change $t_{\gamma}$ and corresponding date, with percent change for US states.} \\
\label{Tallus}
{\bf State} & {\bf $\gamma_1$} & {\bf $t_{\gamma}$} & {\bf Date (2020)} & {\bf $\gamma_2$} & {\bf \% Change}\\
\hline
\endfirsthead
{\bf State} & {\bf $\gamma_1$} & {\bf $t_{\gamma}$} & {\bf Date (2020)} & {\bf $\gamma_2$} & {\bf \% Change}\\
\hline
\endhead
\hline \multicolumn{4}{r}{\textit{Continued ...}} \\
\endfoot
\hline
\endlastfoot
Alaska & 0.018858 & 91.9803 & Apr 22 & 0.0029998 & -84.0922 \\
Alabama & 0.01584 & 129.1083 & May 29 & 0.0028764 & -81.8412 \\
Arkansas & 0.01501 & 107.0161 & May 07 & 0.006258 & -58.3066 \\
Arizona & 0.019334 & 122.9221 & May 23 & 0.005112 & -73.5597 \\
California & 0.014878 & 111.098 & May 11 & 0.0026375 & -82.2724 \\
Colorado & 0.013955 & 115.4642 & May 15 & 0.0010777 & -92.2773 \\
Connecticut & 0.0077278 & 113.8637 & May 14 & 0.0006074 & -92.14 \\
District of Columbia & 0.01744 & 126.0829 & May 26 & 0.0014485 & -91.6943 \\
Delaware & 0.01066 & 132.3503 & Jun 01 & 0.001541 & -85.5436 \\
Florida & 0.015309 & 129.4741 & May 29 & 0.0042771 & -72.0615 \\
Georgia & 0.011839 & 125.6616 & May 26 & 0.0028935 & -75.5597 \\
Hawaii & 0.0092434 & 94.9948 & Apr 25 & 0.0024744 & -73.231 \\
Iowa & 0.0041224 & 119.2705 & May 19 & 0.00037223 & -90.9704 \\
Idaho & 0.014863 & 110.389 & May 10 & 0.0047981 & -67.7173 \\
Illinois & 0.023902 & 164.9266 & Jul 04 & 0.0043251 & -81.9046 \\
Indiana & 0.01642 & 113.6621 & May 14 & 0.0013726 & -91.6409 \\
Kansas & 0.0066308 & 99.5549 & Apr 30 & 0.00028062 & -95.768 \\
Kentucky & 0.028502 & 128.7954 & May 29 & 0.0057361 & -79.8748 \\
Louisiana & 0.026275 & 137.2672 & Jun 06 & 0.0064757 & -75.3538 \\
Massachusetts & 0.0053833 & 117.0038 & May 17 & 0.00089078 & -83.4529 \\
Maryland & 0.0052426 & 124.0183 & May 24 & 0.00098801 & -81.154 \\
Maine & 0.013773 & 120.8261 & May 21 & 0.0020725 & -84.9523 \\
Michigan & 0.022984 & 108.211 & May 08 & 0.00097971 & -95.7374 \\
Minnesota & 0.012611 & 120.2323 & May 20 & 0.0004156 & -96.7044 \\
Missouri & 0.019139 & 115.7427 & May 16 & 0.0019625 & -89.7462 \\
Mississippi & 0.017016 & 133.2993 & Jun 02 & 0.01054 & -38.0591 \\
Montana & 0.010594 & 96.861 & Apr 27 & 0.0031836 & -69.9485 \\
North Carolina & 0.01474 & 110.1662 & May 10 & 0.0023085 & -84.3386 \\
North Dakota & 0.018339 & 131.7428 & Jun 01 & 0.0057475 & -68.6592 \\
Nebraska & 0.012986 & 104.1514 & May 04 & 0.0046664 & -64.065 \\
New Hampshire & 0.014292 & 96.4596 & Apr 26 & 0.031264 & 118.7539 \\
New Jersey & 0.014136 & 76.3807 & Apr 06 & 0.047291 & 234.5559 \\
New Mexico & 0.01693 & 121.5203 & May 22 & 0.0029829 & -82.3808 \\
Nevada & 0.027147 & 110.4525 & May 10 & 0.0027828 & -89.7491 \\
New York & 0.034929 & 112.4836 & May 12 & 0.0039349 & -88.7347 \\
Ohio & 0.010123 & 127.3317 & May 27 & 0.001216 & -87.9875 \\
Oklahoma & 0.024237 & 95.1332 & Apr 25 & 0.002249 & -90.7209 \\
Oregon & 0.01994 & 105.7741 & May 06 & 0.0033566 & -83.1668 \\
Pennsylvania & 0.0042774 & 112.0262 & May 12 & 0.00088788 & -79.2426 \\
Puerto Rico & 0.015866 & 119.5141 & May 20 & 0.0041358 & -73.933 \\
Rhode Island & 0.0074883 & 99.257 & Apr 29 & 0.011599 & 54.897 \\
South Carolina & 0.01325 & 103.3883 & May 03 & 0.0038586 & -70.8792 \\
South Dakota & 0.0064334 & 203.8722 & Aug 12 & 0.0015365 & -76.1174 \\
Tennessee & 0.0061235 & 107.274 & May 07 & 0.0012167 & -80.1301 \\
Texas & 0.011842 & 84.0796 & Apr 14 & 0.0041029 & -65.3542 \\
Utah & 0.0037314 & 110.1407 & May 10 & 0.0011337 & -69.6166 \\
Virginia & 0.017346 & 124.8077 & May 25 & 0.0066082 & -61.9042 \\
Vermont & 0.023989 & 107.7641 & May 08 & 0.0006306 & -97.3713 \\
Washington & 0.023503 & 123.4539 & May 23 & 0.0047135 & -79.9452 \\
Wisconsin & 0.0044732 & 125.8417 & May 26 & 0.00037944 & -91.5175 \\
West Virginia & 0.019036 & 109.8525 & May 10 & 0.0069387 & -63.5491 \\
Wyoming & 0.0015296 & 132.9206 & Jun 02 & 0.00036233 & -76.3115 \\
\end{longtable}
\end{center}

\begin{center}
\begin{longtable}{llllll}
\caption{\normalsize Death rates $\gamma_1$ and $\gamma_2$, day of change $t_{\gamma}$ and corresponding date, with percent change for European countries. * for change indicates country had no deaths before time $t_{\gamma}$.} \\
\label{Talleu}
{\bf Country} & {\bf $\gamma_1$} & {\bf $t_{\gamma}$} & {\bf Date (2020)} & {\bf $\gamma_2$} & {\bf \% Change}\\
\hline
\endfirsthead
{\bf Country} & {\bf $\gamma_1$} & {\bf $t_{\gamma}$} & {\bf Date (2020)} & {\bf $\gamma_2$} & {\bf \% Change}\\
\hline
\endhead
\hline \multicolumn{4}{r}{\textit{Continued ...}} \\
\endfoot
\hline
\endlastfoot
Albania & 0.046842 & 62.1243 & Mar 23 & 0.0080575 & -82.7986 \\
Andorra & 0.020542 & 116.5576 & May 17 & 0.00055115 & -97.3169 \\
Austria & 0.0098244 & 113.6568 & May 14 & 0.00063225 & -93.5645 \\
Belarus & 0.0047253 & 161.9287 & Jul 01 & 0.020364 & 330.9656 \\
Belgium & 0.027217 & 112.0309 & May 12 & 0.0014119 & -94.8126 \\
Bosnia and Herzegovina & 0.01543 & 115.5272 & May 16 & 0.005373 & -65.1794 \\
Bulgaria & 0.0052442 & 131.303 & May 31 & 0.0018924 & -63.9136 \\
Channel Islands & 0.0078607 & 106.7912 & May 07 & 0.00012372 & -98.4261 \\
Croatia & 0.0081174 & 124.9307 & May 25 & 0.0021014 & -74.112 \\
Cyprus & 0.015286 & 85.6981 & Apr 16 & 0.0014654 & -90.4136 \\
Czechia & 0.0084008 & 111.5784 & May 12 & 0.00078631 & -90.6401 \\
Denmark & 0.023793 & 106.2476 & May 06 & 0.00098962 & -95.8406 \\
Estonia & 0.0024383 & 112.7541 & May 13 & 4.3292e-14 & -100 \\
Faroe Islands & 8.2776e-11 & 216.8633 & Aug 25 & 1.9168e-09 & * \\
Finland & 0.0022821 & 112.5707 & May 13 & 3.8419e-05 & -98.3165 \\
France & 0.062387 & 112.4251 & May 12 & 0.0024589 & -96.0586 \\
Germany & 0.010068 & 117.7146 & May 18 & 0.0004936 & -95.0975 \\
Gibraltar & 4.1871e-14 & 216.2532 & Aug 24 & 0.00062895 & * \\
Greece & 0.0095689 & 128.07 & May 28 & 0.0018301 & -80.8742 \\
Hungary & 0.053932 & 121.291 & May 21 & 0.0033608 & -93.7685 \\
Iceland & 0.0034045 & 100.8264 & May 01 & 3.0382e-08 & -99.9991 \\
Ireland & 0.01058 & 110.7097 & May 11 & 0.00041922 & -96.0377 \\
Italy & 0.055973 & 118.6296 & May 19 & 0.0056654 & -89.8783 \\
Kosovo & 0.0021242 & 178.0724 & Jul 17 & 0.0037917 & 78.4974 \\
Latvia & 1.0121e-08 & 79.0803 & Apr 09 & 0.031484 & * \\
Liechtenstein & 0.0044271 & 94.2358 & Apr 24 & 1.9343e-08 & -99.9996 \\
Lithuania & 0.014566 & 131.6754 & Jun 01 & 0.001285 & -91.1785 \\
Luxembourg & 0.0089991 & 117.8828 & May 18 & 0.00082332 & -90.851 \\
Malta & 0.0012141 & 130.8727 & May 31 & 8.8242e-05 & -92.7317 \\
Isle of Man & 0.019654 & 134.6321 & Jun 04 & 0.00042544 & -97.8354 \\
Moldova & 0.01743 & 168.4762 & Jul 07 & 0.0096963 & -44.3701 \\
Monaco & 0.010934 & 103.4746 & May 03 & 4.7518e-14 & -100 \\
Montenegro & 0.0041409 & 199.6697 & Aug 08 & 0.0025586 & -38.2119 \\
Netherlands & 0.026892 & 125.8466 & May 26 & 0.0017659 & -93.4335 \\
North Macedonia & 0.0029871 & 170.4689 & Jul 09 & 0.00082065 & -72.5268 \\
Norway & 0.0014174 & 98.9619 & Apr 29 & 0.00013942 & -90.1642 \\
Poland & 0.016209 & 123.6755 & May 24 & 0.0031553 & -80.5332 \\
Portugal & 0.016931 & 137.2254 & Jun 06 & 0.003978 & -76.5041 \\
Romania & 0.027858 & 123.7185 & May 24 & 0.010543 & -62.1555 \\
Serbia & 0.011069 & 46.1611 & Mar 07 & 0.014431 & 30.3728 \\
Slovakia & 0.0019954 & 113.4444 & May 13 & 0.00013117 & -93.4265 \\
Slovenia & 0.0028753 & 97.7511 & Apr 28 & 0.00018597 & -93.5319 \\
San Marino & 0.12548 & 77.8736 & Apr 08 & 0.008933 & -92.8811 \\
Spain & 0.044385 & 102.3834 & May 02 & 0.0012682 & -97.1427 \\
Sweden & 0.040327 & 123.7356 & May 24 & 0.0051592 & -87.2067 \\
Switzerland & 0.013404 & 105.5814 & May 06 & 0.00035987 & -97.3152 \\
Turkey & 0.0079824 & 124.1928 & May 24 & 0.0039202 & -50.8902 \\
United Kingdom & 0.051358 & 131.4971 & May 31 & 0.0098487 & -80.8233 \\
Ukraine & 0.017525 & 147.2525 & Jun 16 & 0.010609 & -39.4664 \\
\end{longtable}
\end{center}
%

\normalsize


%
%
%



\begin{thebibliography}{10}

{\raggedright
\bibitem{horwitz2020trends}
Leora Horwitz, Simon~A Jones, Robert~J Cerfolio, Fritz Francois, Joseph Greco,
  Bret Rudy, and Christopher~M Petrilli.
\newblock Trends in {Covid-19} risk-adjusted mortality rates.
\newblock {\em {Journal of Hospital Medicine}}, October 2020.

\bibitem{novelcoronavirus}
World~Health Organization.
\newblock Novel coronavirus -- {C}hina.
\newblock Technical report, World Health Organization, 2020.

\bibitem{world2020coronavirus22}
World~Health Organization et~al.
\newblock Coronavirus disease 2019 ({COVID}-19): situation report, 22.
\newblock Technical report, World Health Organization, 2020.

\bibitem{han2020lessons}
Emeline Han, Melisa Mei~Jin Tan, Eva Turk, Devi Sridhar, Gabriel~M Leung, Kenji
  Shibuya, Nima Asgari, Juhwan Oh, Alberto~L Garc{\'\i}a-Basteiro, Johanna
  Hanefeld, et~al.
\newblock Lessons learnt from easing {COVID-19} restrictions: an analysis of
  countries and regions in {Asia Pacific and Europe}.
\newblock {\em The Lancet}, 396(10261):1525--1534, November 2020.

\bibitem{hor2020hope}
Jason Horowitz.
\newblock Hope and worry mingle as countries relax coronavirus lockdowns.
\newblock {\em The New York Times}, May 2020.

\bibitem{get2020as}
Jeffrey Gettleman.
\newblock As virus infections surge, countries end lockdowns.
\newblock {\em The New York Times}, June 2020.

\bibitem{20201020-weekly-epi-update-10}
World~Health Organization et~al.
\newblock {COVID-19} weekly epidemiological update, data as received by {WHO}
  from national authorities, as of {18 October 2020, 10 am CEST}.
\newblock Technical report, World Health Organization, 2020.

\bibitem{crow2020euro}
Michael Crowley and Maggie Astor.
\newblock European nations return to restrictions as virus surges.
\newblock {\em The New York Times}, October 2020.

\bibitem{merv2020it}
Sarah Mervosh and Lucy Tompkins.
\newblock `{I}t has hit us with a vengeance': {V}irus surges again across {the
  United States}.
\newblock {\em The New York Times}, October 2020.

\bibitem{drum2020deathrate}
Kevin Drum.
\newblock If {COVID-19} cases are going up, why is the death rate going down?
\newblock {\em {Mother Jones}}, June 2020.

\bibitem{justice2020deaths}
Lauren Justice.
\newblock {U.S. Coronavirus} cases are rising sharply, but deaths are still
  down.
\newblock {\em New York Times}, July 2020.

\bibitem{campbell2020deathrate}
John Campbell.
\newblock Coronavirus, death rates plummet.
\newblock Podcast, August 2020.

\bibitem{yang2020estimating}
Wan Yang, Sasikiran Kandula, Mary Huynh, Sharon~K Greene, Gretchen Van~Wye,
  Wenhui Li, Hiu~Tai Chan, Emily McGibbon, Alice Yeung, Don Olson, et~al.
\newblock Estimating the infection-fatality risk of {SARS-CoV-2} in {New York
  City} during the spring 2020 pandemic wave: a model-based analysis.
\newblock {\em The Lancet Infectious Diseases}, October 2020.

\bibitem{dennis2020improv}
John~M Dennis, Andrew~P McGovern, Sebastian~J Vollmer, and Bilal~A Mateen.
\newblock Improving survival of critical care patients with {Coronavirus}
  disease 2019 in {England}.
\newblock {\em Critical Care Medicine}, Online First, October 26, 2020, 2020.

\bibitem{khan2020predictive}
ZS~Khan, F~Van~Bussel, and F~Hussain.
\newblock A predictive model for {Covid-19} spread--with application to eight
  {US} states and how to end the pandemic.
\newblock {\em {Epidemiology \& Infection}}, 148:1--40, October 2020.

\bibitem{csse2020data}
Ensheng Dong, Hongru Du, and Lauren Gardner.
\newblock {An interactive web-based dashboard to track COVID-19 in real time}.
\newblock {\em {The Lancet infectious diseases}}, 20(5):533--534, 2020.

\bibitem{owid2020mortality}
Max Roser, Hannah Ritchie, Esteban Ortiz-Ospina, and Joe Hasell.
\newblock Mortality risk of {COVID}-19.
\newblock {\em Our World in Data}, 2020.
\newblock \texttt{https://ourworldindata.org/ mortality-risk-covid}.

\bibitem{salazar2020treatment}
Eric Salazar, Paul~A Christensen, Edward~A Graviss, Duc~T Nguyen, Brian
  Castillo, Jian Chen, Bevin~V Lopez, Todd~N Eagar, Xin Yi, Picheng Zhao,
  et~al.
\newblock Treatment of coronavirus disease 2019 patients with convalescent
  plasma reveals a signal of significantly decreased mortality.
\newblock {\em {The American Journal of Pathology}}, 190(11):2290--2303, 2020.

\bibitem{klopfenstein2020tocilizumab}
Timoth{\'e}e Klopfenstein, Souheil Zayet, Anne Lohse, Jean-Charles Balblanc,
  Julio Badie, Pierre-Yves Royer, Lynda Toko, Chaouki Mezher, Marie Bossert,
  Ana-Maria Bozgan, et~al.
\newblock Tocilizumab therapy reduced intensive care unit admissions and/or
  mortality in {COVID-19} patients.
\newblock {\em {M{\'e}decine et Maladies Infectieuses}}, 50(5):397--400, August
  2020.

\bibitem{biran2020tocilizumab}
Noa Biran, Andrew Ip, Jaeil Ahn, Ronaldo~C Go, Shuqi Wang, Shivam Mathura,
  Brittany~A Sinclaire, Urszula Bednarz, Michael Marafelias, Eric Hansen,
  et~al.
\newblock Tocilizumab among patients with covid-19 in the intensive care unit:
  a multicentre observational study.
\newblock {\em {The Lancet Rheumatology}}, 2(10):e603--e612, 2020.

\bibitem{USGDP2020}
Bureau of~Economic~Analysis.
\newblock Gross domestic product by state, 4th quarter and annual 2019, 2020.
\newblock \texttt{http://www.bea.gov}.

\bibitem{EUGDP2020}
International~Monetary Fund.
\newblock World economic outlook april 2018 edition, gdp nominal per capita --
  international dollar, 2018.
\newblock \texttt{http://www.imf.org}.

\bibitem{USGini2020}
United States~Census Bureau.
\newblock American community survey data, 2020.
\newblock \texttt{http://www.census.gov}.

\bibitem{EUGini2020}
World Bank.
\newblock Gini index (world bank estimate), 2020.
\newblock \texttt{http://www.data.worldbank.org}.

\bibitem{USage2020}
StatsAmerica.
\newblock Median age in 2018, 2020.
\newblock \texttt{http://www.statsamerica.org}.

\bibitem{EUage2020}
Central~Intelligence Agency.
\newblock The world factbook/median age, 2018.
\newblock \texttt{http://www.cia.gov}.

\bibitem{maghbooli2020vitamin}
Zhila Maghbooli, Mohammad~Ali Sahraian, Mehdi Ebrahimi, Marzieh Pazoki, Samira
  Kafan, Hedieh~Moradi Tabriz, Azar Hadadi, Mahnaz Montazeri, Mehrad Nasiri,
  Arash Shirvani, et~al.
\newblock Vitamin {D} sufficiency, a serum 25-hydroxyvitamin {D} at least 30
  ng/ml reduced risk for adverse clinical outcomes in patients with {COVID-19}
  infection.
\newblock {\em {PloS one}}, 15(9):e0239799, 2020.

\bibitem{betsch2020social}
Cornelia Betsch, Lars Korn, Philipp Sprengholz, Lisa Felgendreff, Sarah Eitze,
  Philipp Schmid, and Robert B{\"o}hm.
\newblock Social and behavioral consequences of mask policies during the
  {COVID-19} pandemic.
\newblock {\em {Proceedings of the National Academy of Sciences}},
  117(36):21851--21853, 2020.

\bibitem{haischer2020wearing}
Michael~H Haischer, Rachel Beilfuss, Meggie~Rose Hart, Lauren Opielinski, David
  Wrucke, Gretchen Zirgaitis, Toni~D Uhrich, and Sandra~K Hunter.
\newblock Who is wearing a mask? {G}ender-, age-, and location-related
  differences during the {COVID-19} pandemic.
\newblock {\em {PloS one}}, 15(10):e0240785, 2020.

\bibitem{titus2020mathematical}
Rotich~Kiplimo Titus, Lagat~Robert Cheruiyot, and Choge~Paul Kipkurgat.
\newblock Mathematical modeling of {Covid-19} disease dynamics and analysis of
  intervention strategies.
\newblock {\em {Mathematical Modelling and Applications}}, 5(3):176, 2020.

\bibitem{stvrivzova2020can}
Zuzana St{\v{r}}{\'i}{\v{z}}ov{\'a}, Ji{\v{r}}ina Bart{\u{u}}{\v{n}}kov{\'a},
  Daniel Smr{\v{z}}, et~al.
\newblock Can wearing face masks in public affect transmission route and viral
  load in {COVID-19}?
\newblock {\em {Central European Journal of Public Health}}, 28(2):161--162,
  2020.

\bibitem{gandhi2020masks}
Monica Gandhi, Chris Beyrer, and Eric Goosby.
\newblock Masks do more than protect others during {COVID-19}: reducing the
  inoculum of {SARS-CoV-2} to protect the wearer.
\newblock {\em {Journal of General Internal Medicine}}, 35(10):3063--3066,
  October 2020.

\bibitem{yougov2020masks}
Max Roser, Hannah Ritchie, Esteban Ortiz-Ospina, and Joe Hasell.
\newblock {YouGov COVID-19} behaviour changes tracker: Wearing a face mask when
  in public places.
\newblock {\em YouGov}, 2020.
\newblock \texttt{https://yougov.co.uk/topics/health/articles-reports/
  2020/07/27/face-mask-use-surges-after-becoming-compulsory-sho/}.

\bibitem{leung2020empirical}
Kathy Leung, Yao Pei, Gabriel~M Leung, Tommy~TY Lam, and Joseph~T Wu.
\newblock {Empirical transmission advantage of the D614G mutant strain of
  SARS-CoV-2}.
\newblock {\em {medRxiv}}, 2020.
\newblock 2020.09.22.20199810.

\bibitem{korber2020spike}
Bette Korber, Will Fischer, S~Gnana Gnanakaran, Heyjin Yoon, James Theiler,
  Werner Abfalterer, Brian Foley, Elena~E Giorgi, Tanmoy Bhattacharya,
  Matthew~D Parker, et~al.
\newblock {Spike mutation pipeline reveals the emergence of a more
  transmissible form of SARS-CoV-2}.
\newblock {\em {bioRxiv}}, 2020.
\newblock 2020.04.29.069054.

\bibitem{gomez2020mapping}
Alberto G{\'o}mez-Carballa, Xabier Bello, Jacobo Pardo-Seco, Federico
  Martin{\'o}n-Torres, and Antonio Salas.
\newblock {Mapping genome variation of SARS-CoV-2 worldwide highlights the
  impact of COVID-19 super-spreaders}.
\newblock {\em {Genome Research}}, 30(10):1434--1448, 2020.

\bibitem{mohammad2020higher}
Anwar Mohammad, Eman Alshawaf, Sulaiman~K Marafie, Mohamed Abu-Farha, Jehad
  Abubaker, and Fahd Al-Mulla.
\newblock {Higher binding affinity of Furin to SARS-CoV-2 spike (S) protein
  D614G could be associated with higher SARS-CoV-2 infectivity}.
\newblock {\em {International Journal of Infectious Diseases}}, October 2020.

\bibitem{long2020molecular}
S~Wesley Long, Randall~J Olsen, Paul~A Christensen, David~W Bernard, James~J
  Davis, Maulik Shukla, Marcus Nguyen, Matthew~Ojeda Saavedra, Prasanti
  Yerramilli, Layne Pruitt, et~al.
\newblock {Molecular architecture of early dissemination and massive second
  wave of the SARS-CoV-2 virus in a major metropolitan area}.
\newblock {\em {mBio}}, 11(6), 2020.

\bibitem{becerra2020sars}
Manuel Becerra-Flores and Timothy Cardozo.
\newblock {SARS-CoV-2 viral spike G614 mutation exhibits higher case fatality
  rate}.
\newblock {\em {International Journal of Clinical Practice}}, 2020.
\newblock 74:e13525.

\bibitem{pachetti2020impact}
Maria Pachetti, Bruna Marini, Fabiola Giudici, Francesca Benedetti, Silvia
  Angeletti, Massimo Ciccozzi, Claudio Masciovecchio, Rudy Ippodrino, and
  Davide Zella.
\newblock {Impact of lockdown on Covid-19 case fatality rate and viral
  mutations spread in 7 countries in Europe and North America}.
\newblock {\em {Journal of Translational Medicine}}, 18(1):1--7, 2020.

\bibitem{leffler2020association}
Christopher~T Leffler, Edsel~B Ing, Joseph~D Lykins, Matthew~C Hogan, Craig~A
  McKeown, and Andrzej Grzybowski.
\newblock {Association of country-wide {C}oronavirus mortality with
  demographics, testing, lockdowns, and public wearing of masks. Update August
  4, 2020.}
\newblock {\em {medRxiv}}, 2020.
\newblock 2020.05.22.20109231.

\bibitem{terriau2020impact}
Anthony Terriau, Julien Albertini, Arthur Poirier, and Quentin Le~Bastard.
\newblock {Impact of virus testing on COVID-19 case fatality rate: estimate
  using a fixed-effects model}.
\newblock {\em {medRxiv}}, 2020.
\newblock 2020.04.26.20080531.

\bibitem{jhu2020testing}
Testing Hub.
\newblock Daily state-by-state testing trends.
\newblock {\em Johns Hopkins University of Medicine Coronavirus Resource
  Center}, 2020.
\newblock \texttt{https://coronavirus.jhu.edu/ testing/individual-states}.

\bibitem{promislow2020gero}
Daniel~EL Promislow.
\newblock A geroscience perspective on {COVID-19} mortality.
\newblock {\em The Journals of Gerontology: Series A}, 75:e30--e33, 2020.

\bibitem{venkatesan2020changing}
Priya Venkatesan.
\newblock {The changing demographics of COVID-19}.
\newblock {\em {The Lancet Respiratory Medicine}}, October 2020.

\bibitem{griff2020wuhan}
James Griffiths and Steven Jiang.
\newblock {Wuhan officials have revised the city's coronavirus death toll up by
  50\%}.
\newblock {\em {CNN}}, April 2020.

\bibitem{staff2020update}
Reuters Staff.
\newblock {Argentina's Coronavirus death toll leaps above 20,000 as new data
  added}.
\newblock {\em {Reuters Healthcare \& Pharma}}, October 2020.

\bibitem{hene2020public}
Carl Heneghan and Jason Oke.
\newblock {Public Health England has changed its definition of deaths: here?s
  what it means}.
\newblock {\em {CEBM The Centre for Evidence-Based Medicine}}, August 2020.

\bibitem{nix2020news}
Jamie Nixon.
\newblock {News Release: June 17, 2020 (20-107) Department of Health adjusting
  reporting of COVID-19 related deaths}.
\newblock {\em {Washington State Department of Health}}, June 2020.

\bibitem{blenk2020belg}
Philip Blenkinsop.
\newblock {Belgium revises down COVID-19 deaths just shy of 10,000 mark}.
\newblock {\em {Reuters Healthcare \& Pharma}}, August 2020.
}
\end{thebibliography}
\end{document}